# Effect of electron-beam irradiation on graphene field effect devices

Isaac Childres[1,2], Luis A. Jauregui[2,3], Mike Foxe[4,#], Jifa Tian[1,2], Romaneh Jalilian[1,2,*], Igor Jovanovic[4,#], Yong P. Chen[1,2,3,$]

Abstract: Electron beam exposure is a commonly used tool for fabricating and imaging graphene-based devices. Here we present a study of the effects of electron-beam irradiation on the electronic transport properties of graphene and the operation of graphene field-effect transistors (GFET). Exposure to a 30 keV electron-beam caused negative shifts in the charge-neutral point (CNP) of the GFET, interpreted as due to n-doping in the graphene from the interaction of the energetic electron beam with the substrate. The shift of the CNP is substantially reduced for suspended graphene devices. The electron beam is seen to also decrease the carrier mobilities and minimum conductivity, indicating defects created in the graphene. The findings are valuable for understanding the effects of radiation damage on graphene and for the development of radiation-hard graphene-based electronics.

[1]*Department of Physics, Purdue University, West Lafayette, IN, 47907, USA*
[2]*Birck Nanotechnology Center, Purdue University, West Lafayette, IN, 47907, USA*
[3]*School of Electrical and Computer Engineering, Purdue University, West Lafayette, IN, 47907, USA*
[4]*School of Nuclear Engineering, Purdue University, West Lafayette, IN, 47907, USA*
[#]*Current Address: Department of Mechanical and Nuclear Engineering, The Pennsylvania State University, University Park, PA, 16802, USA*
[*]*Current Address: NaugaNeedles, Louisville, KY, 40299, USA*
[$]*Email: yongchen@purdue.edu*

Graphene has been the focus of much research in material science and nanotechnology due to its unique properties and potentials in device applications. Many reports have been made on graphene's very high electrical conductivity [1,2] at room temperature, and its potential use in next-generation transistors [3], nano-sensors [4], and many other applications.

The effect of e-beam irradiation on graphene and graphene devices is important because of the prevalence of electron beams in both imaging of graphene, e.g. scanning electron microscopy (SEM) and transmission electron microscopy (TEM), and fabrication of graphene devices using e-beam lithography (EBL). In addition, such studies are important to develop radiation-hard graphene-based electronics that can stand up to extreme conditions such as charged particle irradiation in space [5].

Several recent works in the field of energetic particle irradiation of graphene have used positive ions [6-11] or protons [12]. It has been suggested that such irradiations create lattice defects in graphene. There have been studies using energetic electron-beam irradiation to create disorder in carbon nanotubes and graphite [13] and others that focus mostly on the Raman spectroscopy of the electron-beam-induced defects in graphene [14-16]. In this study, we present data on the effect of energetic electron irradiation on the electrical transport properties of single-layer graphene and the operation of graphene field-effect transistors (GFET).

Our graphene samples are fabricated by micromechanical exfoliation [1] of highly ordered pyrolytic graphite (HOPG, ZYA grade, Momentive Performance Materials) onto a p++ (boron-doped, with room temperature resistivity < 0.005 ohm-cm) Si wafer covered with 300 nm of $SiO_2$. Single-layer graphene flakes, typically around 100 $\mu m^2$ in size, are identified using color contrast with an optical microscope [17], and then confirmed with Raman spectroscopy (using a 532 nm excitation laser) [18]. Graphene field-effect devices are subsequently fabricated using EBL. The electrical contacts (5 nm-thick chromium and 65 nm-thick gold) are fabricated by electron-beam evaporation.

A graphene device is placed in a scanning electron microscope (EVO40) under high vacuum ($10^{-6}$ torr). An area of 25μm by 25μm (shown in Figure 1(a) inset as the black-bordered box indicated by the arrow), including the graphene flake on the device, is exposed to the electron beam. The electron beam's kinetic energy is 30 keV, the same energy that is used for our lithography and imaging processes. The beam current ($I_e$) used ranges from 0.15 nA o 0.33 nA. The product of $I_e$ with accumulated exposure time ($T_e$) gives the accumulated irradiation dosage (DOS) (e.g. $T_e$ = 75 s and $I_e$ = 0.15 nA gives DOS = 112.5 e$^-$/nm$^2$). In comparison, the typical exposure used in our lithography process is around 1 e$^-$/nm$^2$. SEM imaging typically exposes samples to at least 100 e$^-$/nm$^2$.

After each successive exposure, the graphene device is removed from the scanning electron microscope, and then room-temperature electrical or Raman measurements are promptly performed. Field-effect electrical measurements with the p-doped Si substrate as the back gate are performed using a probe station filled with argon gas at 1 atm. Raman spectra are taken with a 532 nm excitation laser in an ambient atmosphere.

Results from three gate voltage sweeps (field-effect) measured from a representative device ("A") are shown in Figure 1(a). The conductivity ($\sigma$) is determined by 4-terminal resistance measurements using low-frequency lock-in detection. Initially (before exposure), the device shows a charge-neutral "Dirac" point (CNP, defined as where $\sigma$ is at a minimum [1]) of 16.3 V. The positive CNP is typical in our fabricated devices because of extrinsic hole doping in graphene from, eg., water molecules in the air [19] and resist residues from the lithography process [20]. After the device is exposed to the electron beam with DOS = 112.5 e$^-$/nm$^2$, we observe an appreciable negative shift of the CNP to 4.9 V.

After a larger DOS = 4500 e$^-$/nm$^2$ (accumulated from multiple exposures), the CNP decreases further to −3.8 V, and the slope of the field-effect curve (away from the CNP), related directly to the carrier mobility, decreases significantly in magnitude. The minimum conductivity ($\sigma_{min}$, taken as $\sigma$ at the CNP) also decreases substantially.

Figure 1(b) shows Raman spectra on a similar sample. We observe the appearance of the disorder-induced "D" peak after electron-beam irradiation. This is similar to what was observed previously [14-16], indicating defects created by electron-beam irradiation in graphene.

In this work, we focus on the effect of electron-beam irradiation on electronic transport properties. Figure 2 shows the CNP, mobilities and $\sigma_{min}$ of sample "A" for a series of increasing irradiation dosages.

The electron and hole mobilities ($\mu_e$ and $\mu_h$, respectively) are extracted by examining the slope of the field-effect curve, conductivity ($\sigma$) vs back gate voltage ($V_g$), where $V_g$ is sufficiently far away from the CNP and the curve is in the linear regime, using

$$\mu = (t/\varepsilon) \times (d\sigma/dV_g), \tag{1}$$

where t = 300 nm is the thickness of the SiO$_2$ and $\varepsilon = 3.9 \times \varepsilon_0 = 3.45 \times 10^{-11}$ F/m is the permittivity of the SiO$_2$ [19]. During the first several irradiations, the mobilities decrease sharply from ~5000-6000 cm$^2$/Vs (pre-exposure), then begin to saturate at ~1000 cm$^2$/Vs after ~1000 e$^-$/nm$^2$ of accumulated exposure. We also see $\sigma_{min}$, plotted in units of $e^2/h$ (where $e$ is electron charge and $h$ is Planck's constant), decreases by more than a factor of 2, from ~7 $e^2/h$ before exposure to ~3 $e^2/h$.

The CNP decreases from 17 V before exposure to less than 0 V after extended exposure. We interpret most of the negative shift of the CNP as due to the interaction of the SiO$_2$/Si substrate with energetic electron-beam irradiation. This irradiation generates electron-hole pairs, and the less-mobile holes can get trapped at the SiO$_2$/Si interface to create an effective extra positive bias, attracting electrons in the graphene and resulting in a decreased CNP. This is similar to the negative shift of threshold voltage shift well-known in irradiated metal-oxide-semiconductor field-effect transistors (MOSFETs) [5]. Control studies performed using "simulated" exposure procedures (similar to those used in Figure 2), but without actually turning on the electron-beam, show a much smaller negative shift (by ~5 V, compared to ~20 V of shift with the electron beam on) of the CNP (possibly because the SEM vacuum pumping helps remove surface adsorbates on graphene) and negligible changes in $\mu_e$, $\mu_h$ and $\sigma_{min}$.

To further investigate the influence of the substrate on graphene's CNP under energetic electron-beam irradiation, we have also fabricated suspended graphene devices and irradiated them in the same manner [21]. All field-effect measurements on the suspended devices are 2-terminal. Figure 3(a) shows the CNP decrease by less than 0.16 V after DOS = 112.5 e$^-$/nm$^2$ (compared to a ~12 V shift for our typical substrate-supported device). This confirms the importance of the substrate for the observed CNP shift.

The negative CNP shift we observed was not seen in positive ion irradiation studies where the ion kinetic energy was much lower (e.g. 500 eV) [9].

The substantial drop in the mobilities and the characteristic appearance of the Raman "D" band (also observed in suspended graphene, Figure 3(b)) in the graphene samples after exposure indicate that electron beam irradiation is damaging the graphene lattice structure, creating defects that also scatter the carriers.

In summary, we have observed primarily two effects of electron-beam irradiation on graphene field-effect devices. The CNP of the substrate-supported graphene decreased significantly, indicating a doping of the graphene caused by the interaction of the energetic electron beam and the substrate. Also, the graphene mobility decreased significantly and a "D" peak emerged in the Raman spectra, indicating irradiation-induced defects in graphene. Care should be taken when using SEM and TEM to image and EBL to fabricate graphene devices, as extended exposure could result in a degradation of the graphene device's electrical transport properties. On the other hand, the change caused in the CNP is highly dependent on the interaction with the sample substrate, implying suspended graphene devices would be excellent candidates for use in rad-hard electronics.

*Acknowledgments*: This work has been partially supported by National Science Foundation (ECCS#0833689), Department of Homeland Security (#2009-DN-077-15 ARI036-02) and by the Defense Threat Reduction Agency (HDTRA1-09-1-0047). We also thank Leonid Rohkinson for the access to the SEM. Yong P. Chen also acknowledges support from the Miller Family Endowment and IBM.

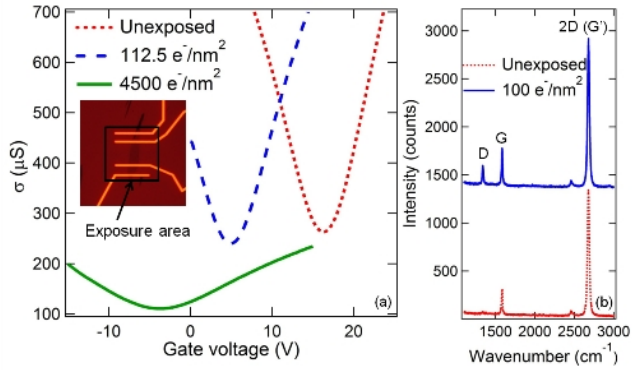

FIG. 1. (a) Measured graphene conductivity as a function of back gate voltage after various doses of electron-beam irradiation for a graphene device on a $SiO_2/Si$ substrate (sample "A"). The source-drain current ($I_{ds}$) used is 100 nA. The inset shows an optical image of the graphene sample measured. (b) Raman spectra before and after irradiation (dosage = 100 $e^-/nm^2$, spectrum offset for clarity) on a similar graphene device (sample "B"). The wavelength of the excitation laser is 532 nm.

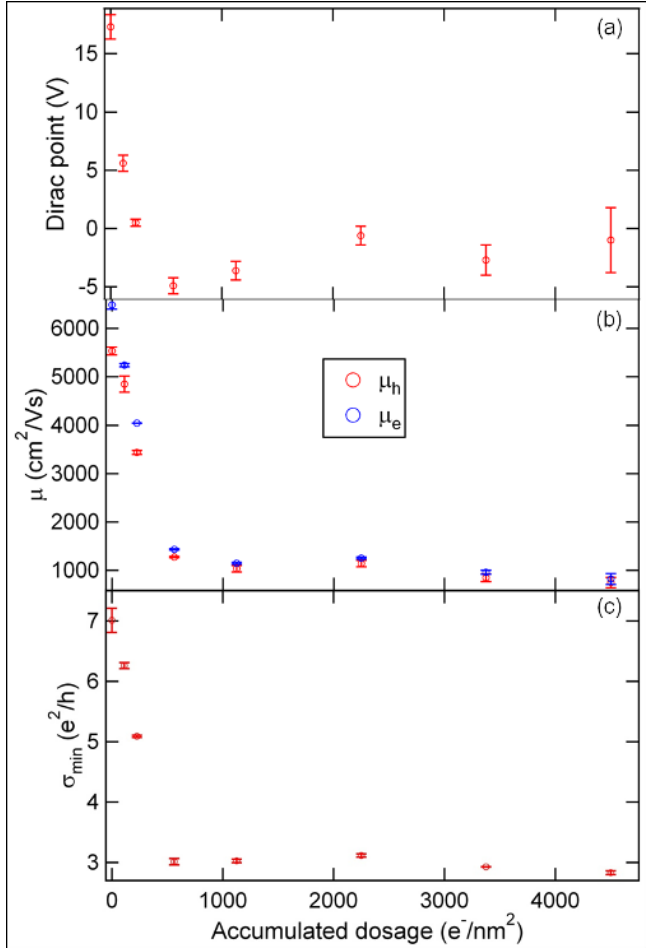

FIG. 2. Charge-neutral "Dirac" point (a), electron and hole field-effect mobilities (b) and minimum conductivity (c) of sample "A" as functions of accumulated electron-beam irradiation dosage. Each data point is the average of two measurements from forward and backward gate voltage sweeps. The error bars reflect the variation between the two sweeps.

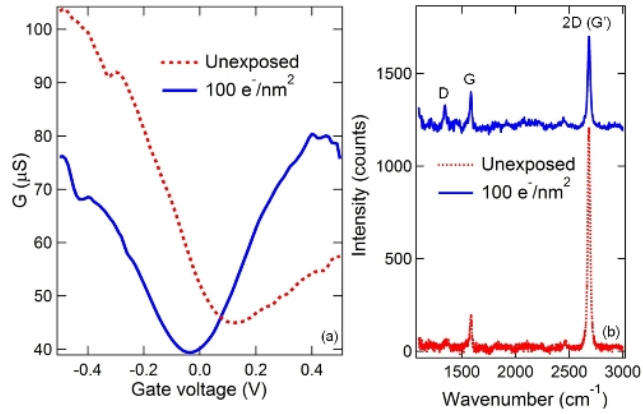

FIG. 3. (a) Measured 2-terminal conductance as a function of back gate voltage before and after electron-beam irradiation (dosage = 112.5 e$^-$/nm$^2$) for a suspended graphene device (sample "C"). The $I_{ds}$ used is 100 nA. (b) Raman spectra (excitation wavelength = 532 nm) taken before and after irradiation (dosage = 100 e$^-$/nm$^2$, spectrum offset for clarity) on a similar suspended graphene device (sample "D").